# GPT-Enabled Cybersecurity Training: A Tailored Approach for Effective Awareness

Nabil Al-Dhamari and Nathan Clarke[0000-0002-3595-3800]

School of Engineering, Computing & Mathematics, University of Plymouth, UK
nclarke@plymouth.ac.uk

**Abstract.** This study explores the limitations of traditional Cybersecurity Awareness and Training (CSAT) programs and proposes an innovative solution using Generative Pre-Trained Transformers (GPT) to address these shortcomings. Traditional approaches lack personalization and adaptability to individual learning styles. To overcome these challenges, the study integrates GPT models to deliver highly tailored and dynamic cybersecurity learning experiences. Leveraging natural language processing capabilities, the proposed approach personalizes training modules based on individual trainee profiles, helping to ensure engagement and effectiveness. An experiment using a GPT model to provide a real-time and adaptive CSAT experience through generating customized training content. The findings have demonstrated a significant improvement over traditional programs, addressing issues of engagement, dynamicity, and relevance. GPT-powered CSAT programs offer a scalable and effective solution to enhance cybersecurity awareness, providing personalized training content that better prepares individuals to mitigate cybersecurity risks in their specific roles within the organization.

## 1 Introduction

The reliance on the Internet and technological advancements has led to a surge in cybersecurity challenges. The escalating digitization of industries has exposed vulnerabilities, making individuals, organizations, and critical infrastructure susceptible to cyber threats. Cybersecurity threats are becoming increasingly sophisticated and dangerous, with a significant rise in cyber-attacks globally, targeting various entities [1,2]. The human factor, encompassing human errors and unintentional misuse, is recognized as a major threat to information systems [3,4]. Indeed, social engineering in particular remains a prevalent form of cyber-attacks, exploiting the human factor through trickery to gain sensitive information [5,6].

Cyber security awareness and training is a vital component of overcoming these human factor issues. However, existing CSAT programs face challenges, notably their lack of dynamism, engagement from individuals, high cost, inflexibility, and difficulty in measuring effectiveness [7-11]. Addressing these challenges, this study proposes leveraging Large Language Models (LLM), specifically GPT models, to create dynamic and personalized CSAT programs. GPT, exemplified by models like GPT-3.5 and GPT-4, utilizes transformer architecture, offering unprecedented capabilities in generating human-like text and performing diverse tasks [12-14]. The study aims to assess the feasibility of using GPT models to tailor security programs, reduce costs, and enhance the learning experience in CSAT.

In the subsequent sections, the paper explores existing literature, proposes, and describes an experiment to explore the feasibility of a GTP-enabled CSAT programme, and discusses the results, implications, and areas for future research. The goal is to provide a comprehensive understanding of how GPT models can revolutionize CSAT programs, addressing the shortcomings observed in traditional approaches [15-16].

## 2    Related Work

The literature review explores current research on (CSAT) programs, employing a keyword search methodology across electronic databases such as IEEE, ResearchGate, Arxiv, and Springer. The keywords include "tailored cyber security awareness programs", "cyber security awareness frameworks", and "artificial intelligence" + "cyber security awareness". Out of 23 relevant papers within the last 5 years, 20 focused on CSAT programs, and 3 on GPT, Large Language Models (LLM), and Deep Learning models.

CSAT research encounters challenges leading to ineffective programs. Various review papers highlight the significance of CSAT for individuals and organizations, along with major limitations. [17] details the impact of cybersecurity awareness, addressing current issues and heightened risks. The review suggests effective delivery methods and a flexible training program to reduce cyberattack severity. [18] employs quantitative research, correlating CSAT programs with employee risk scores. The paper recommends tailored training programs based on employee knowledge levels for improved cybersecurity measures. Additionally, [19] analyze 56 articles on children's cybersecurity awareness, identifying gaps in existing research on risks and training approaches. [20] provide a comprehensive overview of social engineering awareness, emphasizing various factors and suggesting training at all hierarchy levels. Moreover, [21] stress CSAT program importance, proposing a conceptual framework validated through surveys on business managers. Results show prioritizing employee education enhances organizations' security risk management.

Some researchers, like [22], emphasize the need for educational frameworks in cybersecurity campaigns, stressing that knowledge and awareness alone are insufficient to change user behaviors. They advocate for integrating positive cybersecurity behaviors into organizational culture and propose five security factors to enhance CSAT programs. [23] present a novel SETA program based on the behavior change wheel (BCW) framework. In contrast, [24] propose an ADDIE training model using personalized learning theory, offering efficient solutions for individuals and groups. Additionally, [25] focus on developing cybersecurity awareness programs for small and medium enterprises (SMEs), integrating knowledge from academia and practitioners.

[26] address the lack of cybersecurity knowledge among Iranian students with a culturally sensitive ADDIE cybersecurity awareness model. The research emphasizes the integration of cultural influences into cybersecurity education but underscores the need for further cybersecurity training.

Various research efforts contribute innovative approaches to enhance cybersecurity training and awareness. [11] present a cost-benefit analysis framework for CSAT programs, emphasizing the importance of balancing training categories to maintain organizational security. [7] develop a customized SETA program, leveraging the NICE framework and viCyber tool to assess AI-based training effectiveness and promote behavioral change. The application of AI is limited to mapping Knowledge, Skills, and Abilities areas (KSAs) to user competencies, functioning by grouping predefined sets of training data to generate customized training programs. However, it lacks the capability for dynamic content creation, or the provision of finely tailored content based on individual needs. On another hand, [27] focus on specific threats, creating the Cyber Shield game to improve employee cybersecurity awareness in topics such as password management, email attachments, phishing emails, and ransomware threats. [8] propose a dynamic, adaptive cybersecurity training program based on Bloom's taxonomy and STRIDE security models, tailored to trainees' needs, However, it lacks the capability to offer real-time dynamicity, instead establishing a framework for the creation and updating CSAT to align with user needs. Moreover, [28] propose ASURA, a keyword-based educational system providing interactive online training. The system constructs a concept map from the LOD database "DBpedia" using SPARQL querying and employs algorithms to filter irrelevant content. [29] introduce Sifu, a novel cybersecurity platform utilizing gamified AI assessments for software developers, resulting in improved code quality and security.

Examining the intersection of Deep Learning models, Artificial Intelligence, and cybersecurity, researchers have proposed innovative solutions. [12] introduce CyBERT, a specialized BERT model for cybersecurity, by extending the base BERT model's vocabulary to include cybersecurity entities and fine-tuning the model with Masked Language Modeling (MLM) using the cybersecurity corpus. Utilizing Cyber Threat Intelligence (CTI) data and MLM, CyBERT handles tasks like information extraction, attack prediction, and threat classification. The resulting CyBERT model is designed to understand and process cybersecurity-specific textual data for various downstream tasks. However, the CyBERT model requires an extensive domain dataset and the continuous update of its knowledge base to remain relevant. In a broader context, [30] provide a comprehensive overview of Generative AI (GenAI) models, emphasizing their impact on cybersecurity and privacy attacks. The paper outlines potential risks, vulnerabilities, and limitations, addressing ethical and social implications. The researchers discuss potential threats in ChatGPT. In contrast, [13] present a study on the evolution of chatbots, including recent models like GPT-4. The work explores threats, vulnerabilities, and misuse, such as facilitating undetectable zero-day attacks, malware code creation, phishing emails, and ChatGPT's role in cybersecurity.

The studies agree on the importance of CSAT programs, emphasizing dynamic content, engaging delivery methods, and innovative approaches. However, most solutions lack true dynamicity and real-time content creation based on end-user preferences. Additionally, the high cost of keeping CSAT programs up to date remains a significant challenge.

## 3        A Novel GPT CSAT Framework

The primary challenge in current CSAT courses is their lack of dynamism, requiring significant efforts to stay both current and provide appropriate education given an understanding of the learner. GPT, adept at natural conversation, presents a possible solution by dynamically tailoring CSAT content. Success hinges on GPT's ability to adapt using controlled and uncontrolled variables, setting it apart from prior AI attempts. This adaptability reduces costs and effort in customizing CSAT programs, ensuring relevance to evolving cybersecurity knowledge. Adopting a conversational approach personalizes the CSAT experience, allowing users to voice concerns and request tailored content. GPT's human-like conversations could aid knowledge retention through story-based delivery. The study presents an investigation into whether GPT is a viable approach for structuring training courses with a focus on personalized-based learning by considering work experience and job role and subsequently risk individuals might pose. The approach will also utilize organizational-specific security policies to create tailored interactive scenarios.

### 3.1        Structuring the Training Course

Utilizing scenario-based questions improves objective-focused learning, fostering deeper learning [31]. The proposed CSAT program employs a linear approach, presenting real-world scenarios within the user's knowledge domain. GPT's flexibility allows debating options and discussing scenarios in detail, ensuring a comprehensive understanding. According to [14] most employees do not have any deep understanding of their organization's policies beyond the basics, which can pose a risk for the employees and their organizations. With the emphasis on security policies, the program integrates them into the curriculum. Policies drive and control the program, with scenarios and discussions linked back to the policy content. This policy-centered approach ensures technical mastery aligned with organizational security policies.

To address the limitations of traditional CSAT programs, the proposed program leverages GPT's human-like interactivity. Real-time, dynamic conversations, adaptable language, and personalized discussions to enhance the learning experience by replicating the benefits of individual training without prohibitive costs. Recognizing the challenge of inducing behavioral change, the CSAT program prioritizes actionable insights and engagement. GPT's conversational capabilities are utilized to provide authentic support, motivation, and personalized content, fostering a culture of security-first.

### 3.2      Training the GPT Model

GPT acquires knowledge through "model weights" (fine-tuning) or "model inputs" (searching). Fine-tuning suits specialized tasks, while model inputs, providing short-term memory, are preferable for question-and-answer scenarios [32-33]. Training the GPT model is crucial for updating knowledge beyond its last training data, ensuring relevance in rapidly changing fields like cybersecurity. The effectiveness of the proposed CSAT program depends on training the model with organization-specific data, such as security policies.

   GPT models exhibit human-like conversational intelligence by recalling past interactions and maintaining logical coherence. This is achieved through contextual memory processing, which involves storing conversation history, training data, and system guidelines. However, the growing contextual data poses challenges in terms of model request capacity and knowledge processing efficiency. To address these challenges, a "Selective Context" technique is introduced, utilizing Text Embeddings. As illustrated in  Fig. 1, this technique breaks down the flow into phases, each with access to specific contextual data. Essential contextual data, applicable across phases, are prioritized, and summarized information is passed as parameters between phases, reducing the model's memory load.

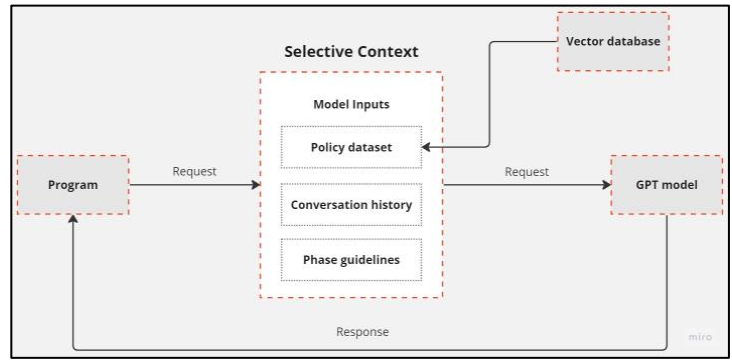

**Fig. 1**. The process of managing the selective context.

   Summarized information, like assessments, is passed between phases as arguments (see Fig. 2), minimizing the model's memory load. Additionally, ongoing conversation parts are automatically summarized and stored, optimizing contextual data management, and allowing for extended conversations. GPT performs separate summarization requests, maintaining information value while minimizing memory usage. These combined techniques enhance the experiment's adaptability and retention of information and personalization settings. The "Selective Context" approach facilitates easy customization by adding or removing phases without sacrificing accuracy or adaptability. The contextual data initially includes pre-loaded policies and high-level instructions. As the program progresses, it will accumulate a summarized version of the conversation history, integrating it with the existing contextual data.

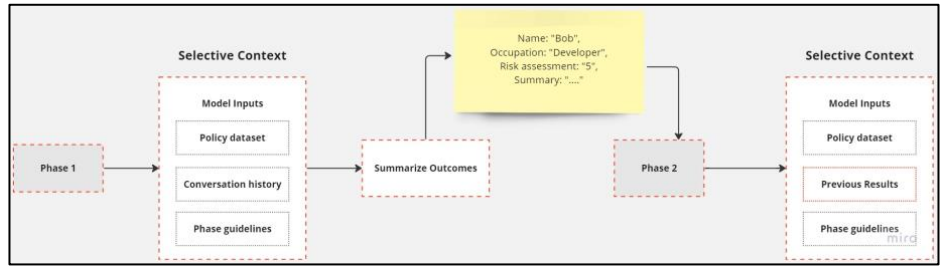

**Fig. 2.** Passing data as arguments between different phases.

### 3.3    Adaptability in CSAT Programs

Adaptability in CSAT programs is crucial for dynamically altering content based on runtime inputs. This feature is pivotal for tailored programs, enabling effective training across diverse roles and contexts. The GPT model's adaptability allows real-time decisions on suitable content and delivery tone, providing a human-like experience. Given varying participant knowledge levels, adaptability ensures program effectiveness for individuals regardless of background, education, or role. To achieve adaptability the model collects critical personal information, conducts knowledge assessments, and generates a risk score for each trainee. This data influences essential contextual data for interactions with the GPT model. The model is instructed to consider these variables in training, demonstrates noticeable shifts in tone, question sophistication, and covered topics. The outcome is a compacted data object containing necessary information for content adaptation. This data object is then passed to the final training phase, enhancing efficiency, reducing API costs, and expanding contextual data for the model's processing.

In summary the core phases of the GPT-based CSAT model are (and illustrated in Fig. 3):

1. Context Setup Phase - Initializes the CSAT program by loading policy datasets and providing system guideline messages to GPT for efficient data retrieval. This phase can be invoked multiple times to modify the model's short-term memory in real-time.
2. Acquaintance Phase - Introduces the CSAT program, emphasizes the security policy, and requests the trainee's details like name, role, and experience. The program summarizes this information, creating a trainee profile.
3. Knowledge Assessment Phase - Builds on the profile to assess the trainee's technical understanding, posing questions about the relevant security policies and practices. The model provides minimal feedback to evaluate the user comprehensively.
4. Risk Assessment Phase - Utilizes collected information, including the trainee's profile, to calculate a risk score manually or through GPT. The model generates a data object encompassing variables such as name, job title, years of experience, risk score, risk summary, and person summary for program adaptation.

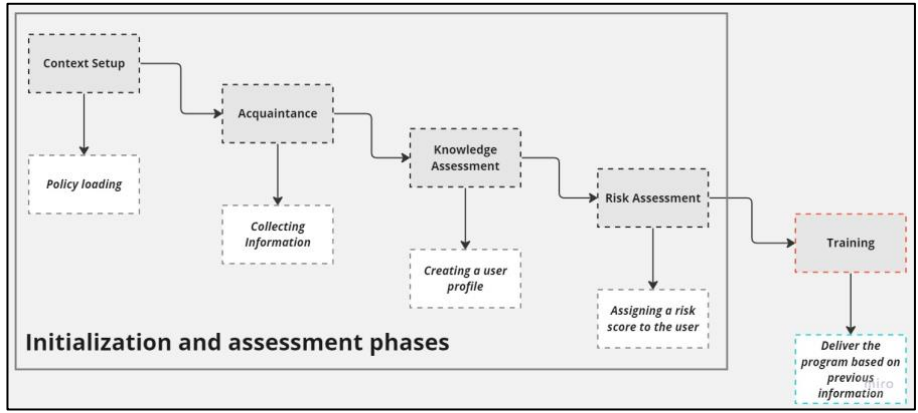

**Fig. 3**. Different phases of the proposed GPT-based CSAT program.

## 4    Experimental Methodology

The proposed GPT-CSAT program centers on policies, program interactivity resembling human-like conversations, and dynamic real-time tailoring. To investigate the degree to which the approach is effective in providing tailored training – both to the organization requirements (via the security policy) and to the individual needs, an experiment was devised in the context of email security. The experiment sought to:

1. Validate the impact of risk scores on program difficulty and structure.
2. Confirm that program personalization aligns with the trainee's job description.
3. Ensure the program remains policy-centered and links questions to policies.
4. Verify the accuracy and relevance of presented information within the CSAT scope.

Table 1 below presents nine fictional personas – developed on the basis that one would expect the nature and diversity of the training to vary given the job role and level of work experience to date. Table 2 defines the different variables that go into the calculation of the *Risk Score*. The *Role Weight* is a manual score linked to a role's responsibilities, gauging the likelihood of it being targeted by attackers. Roles with external email communication or high system privileges receive higher *Role Weights*, indicating an increased risk profile. Conversely, the *Risk Weight* is determined for individual learners during the assessment phase, assigning a score based on their responses and reflecting the risk associated with their current behavior, knowledge, or identified vulnerabilities. These weights collectively offer a thorough evaluation of both role-based and individual-specific security risks in the system. In practice, these weight values are merely used to illustrate the range of different types of learners that exist. The nine permutations were devised with a view to ensuring only one variable changed at a time facilitating the systematic examination of each variable in isolation. By adopting such permutations, the experimental design seeks to discern potential imbalances or biases in the GPT-CSAT responsiveness to individual variables.

Given the ubiquity of email security issues affecting individuals and organizations, especially those with varying knowledge levels, email security training programs were chosen for experiment. The model underwent training using three generic email security policies obtained from the publicly available SANS Institute security policy templates [34]. The GPT-CSAT adaptability relies on unique trainee variables creating a personalized profile. Some variables are inferred during initialization phases, while others are directly collected through trainee responses. Static variables include name, role, and years of experience. Name personalizes communication, role informs about responsibilities and potential targeted attacks, and years of experience gauges the trainee's proficiency in email communication based on tenure.

**Table 1**. Persona profiles utilized in the validation.

| Job Role | Experience (Years) | Risk Score (calculated) |
| --- | --- | --- |
| Social media manager | 1.3 | 7 |
| Social media manager | 2 | 3.5 |
| Social media manager | 1 | 8.5 |
| Chief Finance Officer | 0.5 | 8 |
| IT Vendor Liaison | 0.5 | 4 |
| Customer support specialist | 0.5 | 2 |
| Software QA Engineer | 3.2 | 3.5 |
| Data Analyst | 4.1 | 3.5 |
| Account manager | 1.9 | 3.5 |

**Table 2:** Variables used for manual risk score calculations.

| Variable | Value Range | Description |
| --- | --- | --- |
| Risk weight | 1 - 10 | A risk weight is assigned to the learner based on the results of their initial assessment, the higher the number the more likely this individual is to fall for email attacks based on their answers. |
| Role weight | 1 - 10 | Assigned based on how likely the individual is to be exposed to email attacks in their day-to-day operations. Roles that involve email communications with external entities have a higher role weight. |
| Years of experience | 0.0 - 50.0 | The number of years the learner has worked in the same or similar roles. The more experience they have the less likely they are to fall victim to email attacks. |

Having accumulated comprehensive trainee information, the GPT CSAT approach enters the training phase. The model initiates a new context, incorporating pertinent data from the training set by extracting topics aligned with the trainee's knowledge level using the Text Embeddings API. System guideline messages are given to adjust the program's tone, sophistication, and interaction level based on the trainee's information.

The model receives instructions on scenario-based training, question formulation based on the trainee's role, answer interpretation, and permissible topic deviation. Customizations include presenting email body text examples and email addresses for scenario enrichment.

The *Risk Score* is a core variable that the model uses to aid in delivering the personalization – helping to inform the breadth and depth of knowledge require on a particular topic. The GPT model currently calculates this in a black-box fashion – so the formula being used is unknown. To evaluate whether the model was developing appropriate scores given the input information, the first part of the evaluation focused upon verifying the appropriateness of the scoring. To aid this, a manual approach was also developed based upon the prior art to enable a comparison:

$$Risk\ Score = \frac{(RW\ x\ RX) + y}{(n - 1)}$$

*Where RW = Risk Weight, RX=Role Weight, y=years of experience, n=number of variables utilized*

Given the output from the GPT-CSAT model is textual, a numerical basis for evaluating the performance is challenging and potentially somewhat subjective. As such, it was decided that a simple three-scale metric would be used to determine whether the model had appropriately modified and presented the correct type/level of learning to the recipient persona. This helped to minimize subjectivity that can be introduced using larger scales and provide an initial assessment as to the model's core attributes. Table. 3 outlines these variables with a brief explanation of how criteria and level would be evaluated.

**Table 3**. Variables for measuring the success of experiments.

| Criteria / Ruling | Yes | No | To some extent |
|---|---|---|---|
| Dynamicity based on risk score | Changing the risk score results in a change in the content, tone, mode of delivery, and other aspects of the program. | Changing the risk score does not change anything about the CSAT program. | The CSAT program responds to changes in the risk score, but those changes are not enough to infer dynamicity. |
| Personalization of training based on trainee profile | The program can converse with the trainee using their name and adapt the scenarios based on their occupation with relatable examples. | The program is unable to provide personalized content and treats all the trainees in the same manner. | The program can provide personalized content pertaining to the trainee's name and position, but the scenarios are not related to their occupation. |

| Policy-centered training content | The program does in fact provide policy-centered questions, the follow-up discussions highlight the policy and in general, the program does not deviate from the organization's policy. | The program is easily side-tracked and does not mention the policies anywhere, the questions are random, and the discussions are broad. | The program mentions the policies briefly, but it does not discuss them in greater detail, the program is easily side-tracked and may deviate. |
|---|---|---|---|
| Accuracy of training content | The program provides accurate information about cybersecurity concepts and can mention specific parts of the policies, the model recalls policies by their name and is able to discuss the specifics of each policy. | The program provides inaccurate cybersecurity knowledge or makes false statements about cybersecurity, the program refers to different policies than the ones it was trained on. | The program provides accurate cybersecurity knowledge but has no knowledge of the policies it was trained on which causes it to discuss unknown policies or venture off-topic. |

Utilizing the OpenAI API provides distinct message types allowing varied instructions to the model. This versatility includes *system* messages guiding model behavior, *assistant* messages incorporating conversation history, and *user* messages prompting model responses. API features like Text Embedding aid in breaking down training data, enhancing conversation flow, and preventing topic deviation. Conducted through a Python script using a premium OpenAI API account, experiments spanned GPT3.5-turbo and GPT-4 models.

## 5      Experimental Results

The first part of the experiment sought to evaluate the models' risk-score versus a known manual approach. This would help better understand if the model has appropriately considered the variables given as inputs and determined an appropriate score. As illustrated in Table 4, whilst there is a minor difference in absolute scores (likely based upon the differing formula used), the relative scores (if ordered high to low) follow an identical order. This suggests the GPT-based risk calculation is more than appropriate for estimating the risk profile of individuals.

The experiment was performed on all nine personas. Table 5 presents a summary of the achievements across each persona and Table 6 presents the overall results against the four evaluation criteria. All evaluation criteria were met across all nine personas.

The experiment underscores the importance of multiple message types in the API, clarifying instructions and maintaining a clear separation between *system* and *user* messages. The model's improved tone, explanations, and adherence to predefined instructions are evident. Programmatically controlling the flow and executing varied instructions based on collected variables significantly enhances adaptability and difficulty settings in response to changing risk scores. The GPT-CSAT policy-centered nature is maintained throughout, achieving the defined goals.

**Table 4**. Manual and GPT-based risk score comparison for all permutations.

| Job Title | Years of Experience | Role weight | Risk Weight | GPT Risk Score | Manual Risk Score |
|---|---|---|---|---|---|
| Social media manager | 1.3 | 5 | 2 | 7 | 5.65 |
| Social media manager | 2 | 5 | 1 | 3.5 | 3.5 |
| Social media manager | 1 | 5 | 3.5 | 8.5 | 9.25 |
| Chief Finance Officer | 0.5 | 4 | 4 | 8 | 8.25 |
| IT Vendor Liaison | 0.5 | 5 | 2 | 4 | 5.25 |
| Customer support specialist | 0.5 | 6 | 1 | 2 | 3.25 |
| Software QA Engineer | 3.2 | 3 | 2 | 3.5 | 4.6 |
| Data Analyst | 4.1 | 3 | 1 | 3.5 | 3.55 |
| Account manager | 1.9 | 4 | 2 | 3.5 | 4.95 |

**Table 5**. Analyses of the experiment involving all permutations.

| Permutation No. | Risk Score | Analysis |
|---|---|---|
| 1 | 3.5 | Successful personalization, clear structure, and comprehensive coverage of phishing attacks, strong passwords, and email security policies. Relevant scenario-based questions and interactive discussions improved participant understanding. |
| 2 | 8.5 | Effective personalization, clear structure, and thorough coverage of phishing attacks with tailored multiple-choice questions. Corrections and detailed explanations enhanced participant comprehension. |
| 3 | 7 | Successful personalization, clear structure, and in-depth coverage of various topics with well-structured multiple-choice questions. Detailed corrections and explanations contributed to participant understanding. |
| 4 | 8 | Resounding success with effective personalization, clear structure, and challenging content tailored to the user's role and risk score. Thorough coverage and relevant scenario-based questions improved participant engagement. |
| 5 | 4 | Successful experiment marked by effective personalization, clear structure, and lower-risk scenario tailored with open-ended questions. Thorough coverage, corrections, and discussions enhanced participant education. |

| 6 | 2 | Success achieved through personalized engagement, effective risk recognition, and direct policy references. Thorough coverage, corrections, and a scenario-based question improved participant understanding. |
| 7 | 3.5 | Successful personalization, clear structure, and coverage of social engineering aspects. Open-ended questions encouraged participant explanations. Thorough coverage and corrections contributed to a successful training experience. |
| 8 | 3.5 | Success is marked by personalized content, direct policy references, and practical examples. Thorough coverage, corrections, and scenario-based questions enhanced participant education. |
| 9 | 3.5 | Definite success with effective personalization tailored to the Software QA Engineer role. Thorough coverage, open-ended questions, and corrections improved participant engagement. |

**Table 6**. Measuring the success of experiment

| Goal | Achieved |
|------|----------|
| Dynamicity based on risk score | Yes |
| Personalization of training based on trainee profile | Yes |
| Policy-centered training content | Yes |
| Accuracy of training content | Yes |

Whilst it is not possible to present the complete transcripts that resulted from each of the personas [35], it did seem appropriate to illustrate what the output of GPT-based approach was. Fig. 4, presents the initial welcome, corresponding input from the user regarding role and experience and the response from the GPT model.

```
=====================================================================================
GREETER: Hi there! My name is CyberSentry, and I'm your friendly cybersecurity trainer at ACME. Today, I'm here to talk
to you about the importance of email security. Emails are a vital part of our communication, and it's crucial to keep th
em safe and secure from potential cyber threats. With the increasing number of phishing, malware, and other malicious at
tacks, it's essential to be aware and educated about email security best practices.

    So, let's dive into the world of email security and learn how to keep ourselves and our organization safe. B
ut before we begin, I'd like to get to know you better. May I ask for your name, job title, and level of experience in y
our current role? Once you share that information, we can move forward with our discussion.
=====================================================================================

You: Nabil, social media manager with 1.6 years worth of experience
=====================================================================================
GREETER: Great, Nabil! It's nice to meet you. As a social media manager, you likely handle a lot of communication throug
h emails, making email security even more important for you. Throughout this training program, we will cover various asp
ects of email security, including best practices for protecting sensitive information, identifying and responding to soc
ial engineering attacks, and creating strong passwords.

By the end of this program, you'll have a solid understanding of how to keep your emails secure, protect sensitive compa
ny information, and be vigilant against potential threats. It's crucial to stay updated and informed about email securit
y practices to ensure the safety of our organization and its data.

Now that I know your name, job title, and level of experience, let's get started with our training. Feel free to ask any
questions you may have along the way. Moving on!
=====================================================================================
```

**Fig. 4.** Acquaintance phase through the OpenAI API.

Acquaintance and Risk Assessment Phases show minimal differences across the personas, as they don't require policy knowledge or intricate flow management. Figures 5-6 illustrate the conversation flow. The model actively engages, focuses on policy-related topics during knowledge assessment, and provides detailed evaluations of answers in the risk assessment phase.

**Fig. 5.** Knowledge assessment phase.

**Fig. 6.** Risk assessment phase.

The training phase, pivotal in achieving adaptability, is reliant on information gathered in previous phases (see Fig. 7). The script, interacting with the API, issues system instructions to set tone and persona, along with explicit training directives. Programmatic flow control, driven by user information, allows dynamic structural changes in real-time.

```
"name": "Nabil",
"job title": "Chief Finance Officer",
"years of experience": "0.5",
"risk score": "8",
"summary of risk": "Nabil's job title and years of experience indicate a high level of
    exposure to email-based attacks. However, the lack of knowledge about common types of
    email-based attacks increases the risk. While there is some understanding of indicators
    of phishing emails, there is room for improvement. It is recommended that Nabil
    receives further training and education on email security practices to mitigate the
    identified risks.",
"summary of the person": "The trainee's name is Nabil and they hold the position of Chief
    Finance Officer with 0.5 years of experience. Based on their job title and years of
    experience, it is likely that Nabil has a high level of exposure to email-based attacks
    and is not familiar with the importance of email security practices."
```

**Fig. 7.** Data object from the initialization phases.

Figures 8-11 showcase the dynamic nature of prompt engineering. In this study it was decided that for high-risk trainees, the model is guided through a step-by-step process with multiple-choice questions. Conversely, low-risk scenarios prompt more challenging, open-ended discussions. Low risk here being quantified as an experienced and knowledgeable individual. The power of using multiple message types coupled with controlled program flow becomes evident, allowing tailored interactions based on risk scores.

SYSTEM

When I ask you to ask a scenario-based questions, you will ask me a multiple-choice scenario-based question, such that one option is excellent,
one is good and one is wrong but seemingly correct, and the rest are wrong.

to create the scenario-based question follow these instructions:
1 - Tailor the scenario to fit the job title (social media manager) at ACME and make the scenario tricky.
2 - Tailor the scenario-based on the risk-score (7/10) if the risk-score is low increase the difficulty of the question and vice versa.
3 - Include examples in the scenario, such as email body text or email address to make the scenario more realistic.
4 - Stop at a point where the hypothetical user's action will either deter or cause damage to the organization.
5 - Determine the best course of action and then ask the user what they would have done in this situation.
6 - Offer a few options in random order such that one of them is excellent, one is good and one is wrong but sounds correct and the last one is wrong.
reply only with the question.

**Fig. 8.** Prompt engineering instructions if risk score is higher than the threshold.

```
=====================================================================================
TRAINER: Topic: Phishing Emails

As a Social Media Manager at ACME, you may receive emails from various sources, including potential clients, influencers, and colleagues. However, it's impo
rtant to be cautious as cybercriminals often use phishing emails to trick individuals into revealing sensitive information or downloading malicious attacheme
nts.

Let's consider a scenario:

You receive an email from a popular social media influencer, John Smith, who claims to have a great collaboration opportunity for ACME. The email body state
s that John has a large following and is willing to promote ACME's products/services to his audience. The email also includes a link to a document that supp
osedly contains more details about the collaboration.

Now, here's the scenario-based question:

What would you do in this situation?

A) Click on the link and download the document to learn more about the collaboration opportunity.
B) Reply to the email and ask for more information about the collaboration before taking any further action.
C) Forward the email to your supervisor without clicking on the link or downloading any attachments.
D) Click on the link and provide your login credentials to access the document.

Please choose the option that you think is the best course of action and explain why.
=====================================================================================

You: |
```

**Fig. 9.** Training structure for a high-risk trainee.

```
SYSTEM
To create the scenario-based question follow these instructions:
1 - Tailor the scenario to fit the job title (social media manager) at ACME and make the scenario tricky.
2 - Tailor the scenario-based on the risk-score (3/10) if the risk-score is low increase the difficulty of the
question and vice versa.
3 - Include examples in the scenario, such as email body text and email address to make the scenario
more realistic.
4 - Stop at a point where the hypothetical user's action will either deter or cause damage to the
organization.
5 - Determine the best course of action and then ask the user what they would have done in this
situation.

When I ask you to ask a scenario-based questions, you will ask open-ended questions using one or
more of the following examples:
Example 1 - a scenario where the attacker knows private information about a business meeting that
took place a day before and tricks the user by mentioning details of a meeting with the impersonated
person.
Example 2 - a scenario where the source email looks legitimate by replacing unicode with ascii
characters to fool the user.
Example 3 - a scenario where the user is contacted by the compliance team and urged to download a
document and fill in the information as part of the security compliance operation.
Example 4 - a scenario where the attacker sends a SMS from the impersonated person's phone
number and follows up with a phishing email mentioning the SMS.
```

**Fig. 10.** Prompt engineering instructions if risk score is lower than the threshold.

```
=====================================================================================
TRAINER: Sure! Let's start with the topic of phishing emails. As a Social Media Manager at ACME, you may receive emails from various sources, including pote
ntial clients, partners, or even fans of your company's social media accounts. It's important to be vigilant and cautious when opening and responding to ema
ils to protect sensitive information and prevent unauthorized access to company systems.

Now, let's consider a scenario. Imagine you receive an email from a sender claiming to be a popular social media influencer who wants to collaborate with yo
ur company. The email looks legitimate, with the influencer's name and photo in the signature. The email body mentions that they have been following your co
mpany's social media accounts and are impressed with your work. They propose a collaboration opportunity and ask you to provide your personal email address
and phone number for further communication.

Based on this scenario, here's the question: What would you do in this situation? Would you provide your personal email address and phone number to the send
er, or would you take a different course of action?
=====================================================================================

You: |
```

**Fig. 11.** Training structure with opening ended questions for low-risk trainees.

# 6    Discussion

The GPT-based CSAT program, tailored to individual interests and profiles, has demonstrated effectiveness and potential cost and time savings in developing organizational CSAT programs compared to traditional methods. The chat-style approach enhances usability and flexibility, adapting to individual questions and requirements. However, a significant limitation arises from the inherent variability of GPT-based models in interaction with learners. Even with identical input criteria such as role and work experience, the nature of interaction can differ, introducing uncertainty in performance consistency and effectiveness across all scenarios.

While the evaluation against predefined personas was successful in this study, the consistency of such success in practical applications remains uncertain due to the nascent state of GPT model science. Future research is crucial to establish the reliability and robustness of these models. One alternative strategy suggested is the development of a second independent model to validate and verify outputs, similar to the concept of Generative Adversarial Networks (GANs), which could enhance the overall performance and trustworthiness of GPT-based systems.

The study acknowledges the limitations of using deterministic metrics for evaluation and suggests adopting a more empirical approach. This includes exploring text complexity measures beyond readability to assess subject knowledge complexity and ensuring tailored training delivery based on individual competency levels. Integrating user behavioral data into the GPT model presents an intriguing extension to enhance accuracy and personalization. By adjusting risk scores and weights based on user behavior, the model could provide more personalized and targeted learning experiences, catering to individual strengths and weaknesses.

In summary, while the GPT-based CSAT program shows promise, addressing the variability in interaction and enhancing reliability through alternative validation strategies and user behavioral data integration are critical next steps. Further research and development are needed to optimize the performance and applicability of GPT-based models in diverse learning contexts. These efforts will contribute to advancing the science and practical implementation of AI-driven CSAT programs with enhanced effectiveness and personalized learning experiences.

## 7      Conclusion

In the realm of today's interconnected digital world, cybersecurity plays a pivotal role. Our exploration underscores the significance of the human factor, leading to the development of a dynamic and personalized GPT Cybersecurity Training and Awareness (CSAT) framework. The approach aims to provide continuous, affordable, and high-quality cybersecurity education to individuals and organizations.

OpenAI offers a compelling solution for leveraging and refining cutting-edge Generative AI models that have been extensively pre-trained on a large scale. However, alternative methods also exist for training localized models. This study has concentrated on GPT as it represents the forefront of generative AI technology at the time of writing, yet it encourages further exploration and experimentation with alternative models in future research endeavors.

Future research will focus on refining the data inputs into the model, exploring the incorporation of high-quality, user-specific data, such as external risk score calculations and user behavioral analytics. A participant-based study will also be explored to better understand attitudes and opinions of the approach.